\documentclass{article}%
\usepackage[top=3cm, bottom=3cm, left=3cm, right=3cm]{geometry}
\usepackage{amsmath}
\usepackage{amsfonts}
\usepackage{amssymb}
\usepackage{graphicx}%
\begin{document}

\title{Propagator poles and an emergent stable state below threshold: general
discussion and the $E(38)$ state}
\author{Francesco Giacosa and Thomas Wolkanowski\\\emph{Institut f\"{u}r Theoretische Physik, Johann Wolfgang Goethe -
Universit\"{a}t}\\\emph{Max von Laue--Str. 1, D-60438 Frankfurt am Main, Germany}}
\maketitle

\begin{abstract}
In the framework of a simple quantum field theory describing the decay of a
scalar state into two (pseudo)scalar ones we study the pole(s) motion(s) of
its propagator: besides the expected pole on the second Riemann sheet, we find
-- for a large enough coupling constant -- a second, additional pole on the
first Riemann sheet below threshold, which corresponds to a stable state. We
then perform a numerical study for a hadronic system in which a scalar
particle couples to pions. We investigate under which conditions a stable
state below the two-pion threshold can emerge. In particular, we study the
case in which this stable state has a mass of $38$ MeV, which corresponds to
the recently claimed novel scalar state $E(38)$. Moreover, we also show that
the resonance $f_{0}(500)$ and the stable state $E(38)$ could be two different
manifestations of the same `object'. Finally, we also estimate the order of
magnitude of its coupling to photons.

\end{abstract}

In this Letter we study the propagator, its poles and the spectral function of
a scalar state $S$ which interacts with a (pseudo)scalar field $\varphi$ as
described in the following Lagrangian \cite{veltman,lupo}:
\begin{equation}
\mathcal{L}=\frac{1}{2}(\partial_{\mu}S)^{2}-\frac{1}{2}M_{0}^{2}S^{2}%
+\frac{1}{2}(\partial_{\mu}\varphi)^{2}-\frac{1}{2}m^{2}\varphi^{2}%
+gS\varphi^{2}\text{ .} \label{lag}%
\end{equation}
The propagator of the field $S$ is dressed by loops of $\varphi$-particles.
The contribution of one loop is evaluated using the standard Feynman rules,%

\begin{equation}
\Sigma(p^{2})=-i\int\frac{\text{d}^{4}q}{(2\pi)^{4}}\frac{\phi^{2}(q)}{\left[
\left(  \frac{p}{2}+q\right)  ^{2}-m^{2}+i\varepsilon\right]  \left[  \left(
\frac{p}{2}-q\right)  ^{2}-m^{2}+i\varepsilon\right]  }\text{ ,}%
\end{equation}
where the function $\phi(q)$ has been introduced to regularize the integral.
(Formally, it can be included at the Lagrangian level by rendering it
nonlocal, see Refs.
\cite{nl,efimov,Burdanov:1996uw,Faessler:2003yf,GiacosaGutsche} and references
therein). Here we adopt the choice $\phi(q)=\tilde{\phi}(\mathbf{q}%
^{2})=\left(  1+\mathbf{q}^{2}/\Lambda^{2}\right)  ^{-1}$ \cite{lupo}, where
$\Lambda$ sets the energy scale at which the loop contribution gets
suppressed. It should be stressed that the precise form of the cutoff function
affects the results only marginally, as long as it falls off sufficiently fast
to guarantee convergence.

Upon resummation of the one-loop contributions, the propagator of $S$ takes
the form%
\begin{equation}
G_{S}(p^{2})=\left[  p^{2}-M_{0}^{2}+(\sqrt{2}g)^{2}\Sigma(p^{2}%
)+i\varepsilon\right]  ^{-1}\text{.}\label{prop}%
\end{equation}
The spectral function of the state $S$ is obtained as the imaginary part of
the propagator as \cite{lupo,salam,Matthews:1959sy,achasov}
\begin{equation}
d_{S}(x=\sqrt{p^{2}})=\frac{2x}{\pi}\left\vert \lim_{\varepsilon
\rightarrow0^{+}}\operatorname{Im}[G_{S}(x^{2})]\right\vert \text{ .}%
\end{equation}
As a consequence of the K\"{a}ll\'{e}n--Lehmann representation, the
normalization $\int_{0}^{\infty}d_{S}(x)\text{d}x=1$ holds.

We first describe the general qualitative features of our results, which are
independent on numerical details. We assume that the (bare) mass $M_{0}$ lies
well above the threshold $2m,$ thus implying that the decay into
$\varphi\varphi$ is kinematically allowed. As soon as the coupling constant
$g$ does not vanish, the propagator (\ref{prop}) is a multivalued analytic
function for the complex variable $z=\sqrt{p^{2}}$. The resonance
corresponding to the field $S$ is related to a pole on the (unphysical) second
Riemann sheet. We denote this pole as
\begin{equation}
z_{\text{II}}=M_{pole}-i\Gamma_{pole}/2\text{ }, \label{zii}%
\end{equation}
where $M_{pole}$ is usually referred to as the pole mass of the resonance $S.$

Clearly, for $g\rightarrow0$ one has $M_{pole}\rightarrow M_{0}$ and
$\Gamma_{pole}\rightarrow0$. When varying the coupling constant $g$ a
trajectory in the complex plane is generated, see the dashed line in Fig. 1
for a numerical example and the discussion below. Note, for small values of
$g$ the decay width $\Gamma_{pole}$ obtained from Eq. (\ref{zii}) is
reproduced very well by the tree-level result $\Gamma_{S\varphi\varphi
}(M_{pole})$, where the tree-level decay function $\Gamma_{S\varphi\varphi
}(x)$ reads $\Gamma_{S\varphi\varphi}(x)=\frac{\sqrt{x^{2}/4-m^{2}}}{8\pi
x^{2}}(\sqrt{2}g)^{2}\tilde{\phi}^{2}(x^{2}/4-m^{2})$. In this regime the
spectral function $d_{S}(x)$ is well approximated by a Breit--Wigner function,
see Fig. 2 (solid curve) for a typical Breit--Wigner shaped resonance.
However, for increasing coupling (even sizable) distortions arise \cite{lupo}.
Note, departures from Breit--Wigner also imply a non-exponential decay when
the time evolution of the unstable state is studied \cite{zenoqft,Giacosa2011}%
. This is the quantum field theoretical counterpart of a general quantum
phenomenon, see for example Refs. \cite{ghirardi,facchiprl}.

Besides the pole (\ref{zii}), other poles may exist on the II-Riemann sheet,
which however have practically no influence on the determination of the
properties of the resonance $S$. Interestingly, for moderate values of the
coupling constant $g$ there is a pole on the real axis below threshold. This
pole, although not important for the phenomenology of the resonance, moves for
increasing $g$ along the real axis towards the threshold value $2m$. Then, for
the coupling constant $g$ exceeding a critical value $g_{\ast}$, this pole
disappears from the II-Riemann sheet, but appears on the real axis below
threshold on the I-Riemann sheet. We denote this newly emerged pole on the
first sheet as
\begin{equation}
z_{\text{I}}=\tilde{M}-i\varepsilon\text{ with }\tilde{M}<2m\text{ (for }g\geq
g_{\ast}\text{ only).} \label{zi}%
\end{equation}
The pole $z_{\text{I}}$ changes considerably the physical properties of the
system: namely, it corresponds to a stable state below threshold. Thus, for
$g\geq g_{\ast}$ there are two relevant poles on the Riemann manifold: the
standard resonance pole on the second sheet of Eq. (\ref{zii}) and the
additional `stable state' pole from Eq. (\ref{zi}) on the first sheet. The
spectral function in the presence of the additional pole below threshold takes
the form
\begin{equation}
d_{S}(x)=Z\delta(x-\tilde{M})+d_{a.t.}(x)\text{ ,} \label{dbt}%
\end{equation}
where $d_{a.t.}(x)$ (a.t. stands for above threshold) vanishes for $x<2m.$ The
quantity $Z$ is nonzero for $g\geq g_{\ast}$ and its explicit form is given
by
\begin{equation}
Z=\left[  1+(\sqrt{2}g)^{2}\left(  \frac{\partial\Sigma(x^{2})}{\partial
x^{2}}\right)  _{x^{2}=\tilde{M}^{2}}\right]  ^{-1}\text{ }.
\end{equation}
The full spectral function $d_{S}(x)$ is still normalized to $1$, implying
that $Z+\int_{2m}^{\infty}d_{a.t.}(x)\text{d}x=1$.

The emergence of additional poles was described in the literature (usually
connected to light scalar mesons), e.g. Refs.
\cite{dullemond,tornqvist,boglione,ruppbugg}. As described in detail in Ref.
\cite{ruppbugg}, besides a massive `seed' pole, a second pole on the second
Riemann sheet with a smaller real part exists. The movement of this pole in
the complex plane is peculiar: for small values of (their) coupling constant,
its imaginary part is very large. Then, by increasing the coupling, both the
imaginary and the real part of this additional pole become smaller.
Eventually, for a coupling large enough, a stable state below threshold is
obtained. It should be stressed that the pole we obtain out of our Lagrangian
(\ref{lag}) has very different properties: namely, as described above, it is
situated on the real axis of the second sheet for small and intermediate
values of the coupling, then vanishes for a coupling large enough by slipping
into the branch cut and appears on the real axis of the first sheet.%

\begin{figure}
[ptb]
\begin{center}
\includegraphics[
height=2.9724in,
width=4.7651in
]%
{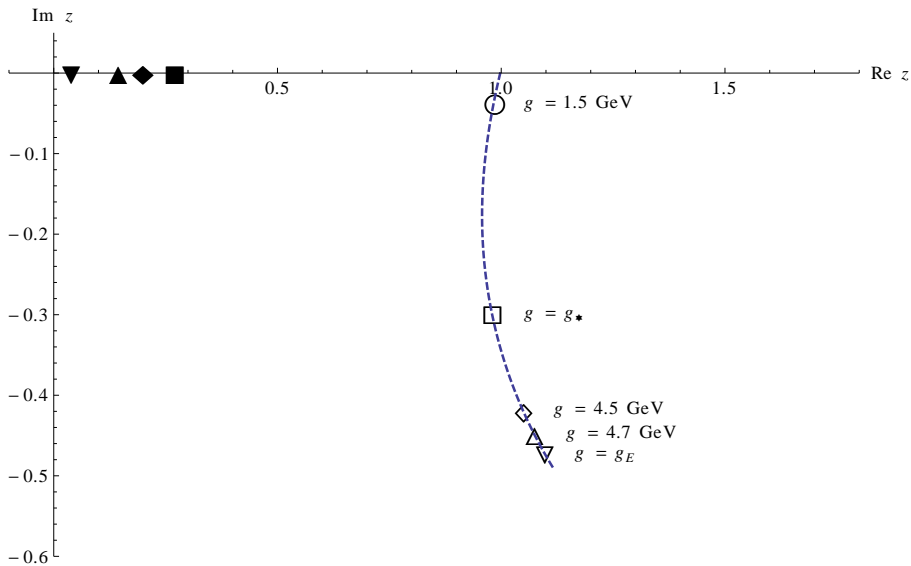}%
\caption{The trajectory (dashed line) of the second-sheet pole $z_{\text{II}}$
(Eq. (\ref{zii})) obtained by varying the coupling constant $g$ is shown for
the numerical values in Eq. (\ref{num}). For some examples (white markers
along the dashed line) the value of $g$ is reported. When existent, we have
also shown -- using the same form of the markers -- the first-sheet pole
$z_{\text{I}}$ (Eq. (\ref{zi})) lying on the real axis below threshold. For
instance, in the case $g=g_{E}$ the pole $z_{\text{I}}$ corresponds to a
stable state with the mass $\tilde{M}=38\ \text{MeV.}$ }%
\label{fig1}%
\end{center}
\end{figure}

We now turn to some numerical calculations in order to show some explicit
examples of the described properties. For illustrative purposes we use the
following numerical values, which are typical in hadronic theories
\cite{amslerrev,klempt}:
\begin{equation}
M_{0}=1\text{ GeV , }\Lambda=1.5\text{ GeV , }m=m_{\pi^{0}}=0.135\text{ GeV .}
\label{num}%
\end{equation}
Thus, the threshold $2m$ corresponds to the minimal energy threshold of QCD,
namely that of two pions. Note, the choice of $M_{0}=1$ GeV in Eq. (\ref{num})
is not linked to a resonance listed in the Particle Data Group (PDG)
\cite{PDG}. It corresponds to a hypothetical resonance with a mass of about 1
GeV which couples strongly to pions. This resonance is supposed to be a
hadronic state made of quarks and gluons (eventually a broad tetraquark
state), but in the present framework it serves just as a tool to test our approach.

When varying the value of the coupling constant $g$ from small to large
values, the position of the resonance pole $z_{\text{II}}$ defines a
trajectory on the II-Riemann sheet, which is depicted in Fig. 1 (dashed line).
Moreover, for $g\geq g_{\ast}=3.739\ \text{GeV}$ the first-sheet (alias
bound-state) pole $z_{\text{I}}=\tilde{M}-i\varepsilon$ is also present and
its position for some values of $g$ is shown.

In particular, we have shown in Fig. 1 also the case $g=g_{E}%
=4.877\ \text{GeV}$ $>g_{\ast}$ for which the mass of the stable state on the
first sheet reads $z_{\text{I}}=\tilde{M}-i\varepsilon=(38-i\varepsilon)$ MeV.
Quite interestingly, in Ref. \cite{e38first} E. van Beveren and G. Rupp
discuss the possible existence of a new boson with a mass of $38$ MeV, called
$E(38),$ by a reanalysis of BABAR data \cite{babar} for the process
$e^{+}e^{-}\rightarrow\Upsilon(1,2^{3}S_{1})\pi^{+}\pi^{-}\rightarrow
l^{+}l^{-}\pi^{+}\pi^{-}$ where $\Upsilon$ stands for a bottonium state and
$l$ for a lepton. Namely, they argue that a light bosonic state with a mass of
$38$ MeV is emitted from the decay of the bottonium, resulting in a missing
mass for the process described above. This hypothetical boson with a mass of
$38$ MeV then further decays into photons. In Ref. \cite{e38material} the same
authors discuss the data of the COMPASS experiment
\cite{compass,Schluter:2011dt}, which allegedly add evidence in favour of the
existence of $E(38)$ and confirms the decay $E(38)\rightarrow\gamma\gamma$.
This results have been disputed in a comment by the COMPASS collaboration
\cite{compasscomment}; later on, E. van Beveren and G. Rupp replied to the
comment \cite{e38reply}. Recently, experimental results from the JINR
Nuclotron from DUBNA claiming the verification of the existence of the $E(38)$
state have been reported in Ref. \cite{dubna}, in which (utterly different)
nuclear reactions involving two photons in the final state have been studied
(presently the work of Dubna is under revision by the authors).

It must be stressed that our main interest is the theoretical description of
the emergence of an additional pole below threshold, independently on the
existence of the putative state $E(38)$; moreover, the issue about the
existence of $E(38)$ is only at its first stage and confirmation is needed. It
could well be that this state is not existent. Nevertheless, due to the fact
that within our simple quantum field theoretical approach we can naturally
accommodate a pole below threshold, it is amusing and interesting to
investigate this numerical example in more details. We thus show in Fig. 2 the
spectral function of the state $S$ for two values of $g$. For the small value
$g=1.5$ GeV there is no pole on the first Riemann sheet and the spectral
function has a typical $70$ MeV broad Breit--Wigner form centered at about
$M_{0}=1$ GeV. For the case $g=g_{E}$ the pole $z_{\text{I}}$ on the first
sheet is present. The spectral function has the form of Eq. (\ref{dbt}) with
$Z=0.28.$ Below threshold there is a $\delta$-type peak for $\tilde{M}=38$
MeV, above threshold there is a very broad and flat enhancement centered at
about $1$ GeV. It is clear that such a structure would be very hard to detect:
its influence on $\varphi\varphi$ (alias pion-pion) scattering is expected to
be very small, thus also explaining why this hypothetical broad state could
not be observed experimentally. Indeed, the coupling of $S$ to pions is large
but not unrealistic: $g_{\pi}=g_{E}/\sqrt{3}=2.816$ GeV, where $\sqrt{3}$
takes into account isospin symmetry, not included in our Lagrangian. A
possibility could be a broad tetraquark state, for which the coupling constant
is of the same order \cite{tq}.%
\begin{figure}
[ptb]
\begin{center}
\includegraphics[
height=2.5598in,
width=4.1831in
]%
{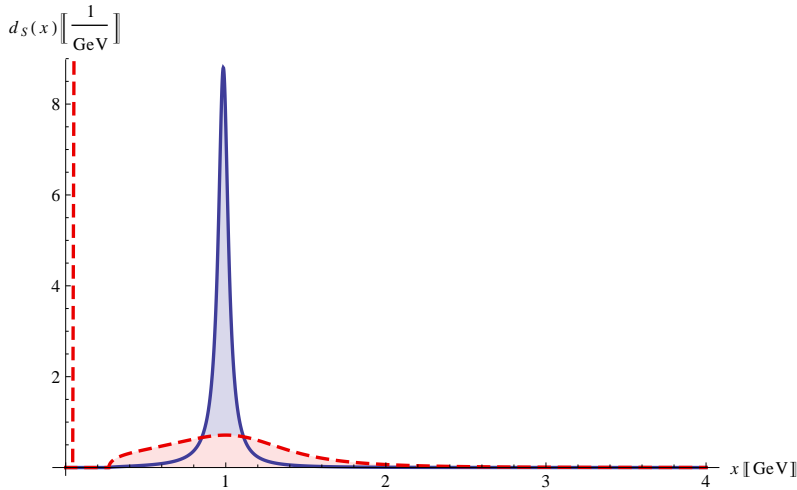}%
\caption{The spectral function of the state $S$ for two different values of
the coupling constant is shown for the numerical values in Eq. (\ref{num}).
The solid (blue) curve corresponds to $g=1.5\ $GeV: the second-sheet pole is
$z_{\text{II}}=(0.985-i0.038)$ GeV. A typical Breit--Wigner form is visible.
The dashed (red) curve corresponds to $g=g_{E},$ for which the second-sheet
pole $z_{\text{II}}=(1.096-i0.472)$ GeV lies deep in the lower half plane.
Being $g_{E}>g_{\ast},$ there is also the first-sheet pole $z_{\text{I}%
}=\left(  0.038\ -i\varepsilon\right)  $ GeV. The spectral function shows a
stable $\delta$-peak for $0.038$ GeV and a broad enhancement above threshold.}%
\end{center}
\end{figure}

A quite interesting and remarkable result follows when testing other values
for $M_{0}.$ Namely, one of the most famous mesonic resonances is the lightest
scalar meson $f_{0}(500).$ In the new version of the PDG the pole of this
resonance is estimated as $(0.40$-$0.55)-i(0.20$-$0.35)$ GeV \cite{PDG}. It is
then natural to ask if we could assign the $f_{0}(500)$ resonance to our state
$S$ and at the same time describe the stable state $E(38)$. Indeed, by simply
changing the mass to $M_{0}=0.45$ GeV we obtain the first-sheet pole
$z_{\text{I}}=(0.038-i\varepsilon)$ GeV for $g=g_{E}=2.189$ GeV. The
second-sheet pole reads $z_{\text{II}}=(0.427-i0.371)$ GeV, which is
\emph{remarkably} close to the range given above. We summarize the results of
this study in Fig. 3, in which both the $z_{\text{II}}$ pole trajectory and
the spectral function for $g=g_{E}$ are shown. It is therefore conceptually
possible to link the states $f_{0}(500)$ and $E(38).$ It is worth to mention
that there are indeed phenomenological models in which the state $f_{0}(500)$
is interpreted as a tetraquark state and where the theory reduces to a
Lagrangian of the type (\ref{lag}) when only the pion-pion channel is studied
\cite{tq,maiani,Hooft:2008we}. Surely, more advanced investigations in this
direction should go beyond our simple model, like testing different cutoff
functions and taking into account other interaction forms, e.g. some that
include derivatives \cite{lupoder}. Most important, one should also include
chiral symmetry. Nevertheless, the possibility to describe $f_{0}(500)$ and
$E(38)$ in an unified framework is intriguing.%

\begin{figure}
[ptb]
\begin{center}
\includegraphics[
height=1.9432in,
width=6.4515in
]%
{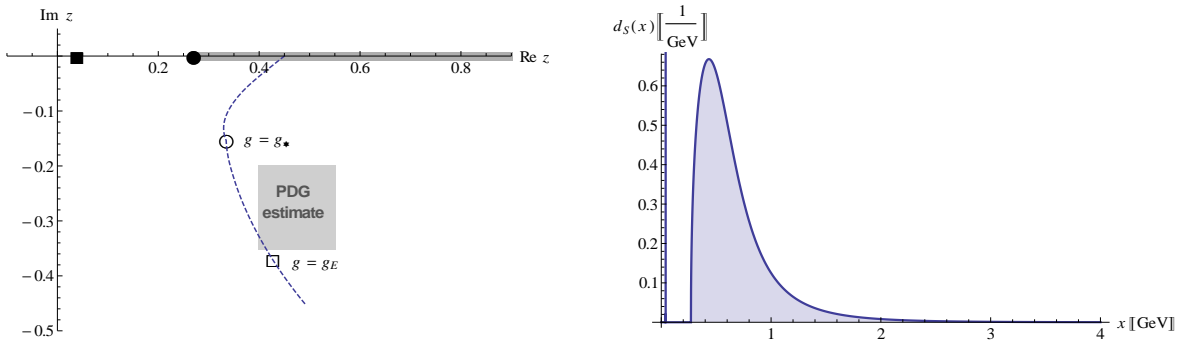}%
\caption{Left: $z_{\text{II}}$ pole trajectory for $M_{0}=0.45$ GeV. The
critical value is here given by $g_{\ast}=1.399$ GeV. For $g=g_{E}=2.189$ GeV
the first-sheet pole $z_{\text{I}}=(0.038-i\varepsilon)$ GeV is obtained. In
the shadowed rectangular the ranges of the PDG estimate for the $f_{0}(500)$
pole is reported: it is visible that the $z_{\text{II }}$pole (white square)
lies very close to it. Right: spectral function $d_{S}(x)$ for the same
numerical values. Here $Z=0.68,$ showing that the pole below threshold gives
the dominant contribution to the spectral function.}%
\end{center}
\end{figure}

An important aspect which needs a more detailed discussion is the validity of
the employed one-loop approximation for the purposes described in this work,
i.e. the emergence of a pole below threshold. It should be noted that the used
one-loop approximation has been resummed, that is a class of diagrams (the
one-loop of $\varphi$-fields) has been considered up to order infinity. One
should investigate the quantitative role that (resummed) higher order diagrams
have, the first of which is the self-energy diagram where a particle of the
$S$-type is exchanged by the two $\varphi$-particles circulating in the loop.
This two-loop calculation represents a (hard) task for the future. In the
Appendix we describe a simpler calculation which helps to shed light on the
role of higher order processes: the vertex correction for the decay process.
In this way it is possible to understand when the dimensionful coupling
constant $g$ can be regarded as small or not. We find that, for the numerical
examples described above, the role of higher order diagrams is negligible (at
least as a first approximation) also for values of $g$ exceeding the critical
value $g_{\ast}$ necessary for the emergence of the stable pole below
threshold. Thus, the inclusion of higher order contributions, while surely
important for a more quantitative analysis, is not expected to change the
qualitative picture described here.

As a last subject of this work we describe the coupling to photons. To this
end we add to the Lagrangian the following interaction term:%
\begin{equation}
\mathcal{L}_{\gamma\gamma}=c_{\gamma\gamma}SF_{\mu\nu}F^{\mu\nu}\text{ },
\end{equation}
where $F^{\mu\nu}=\partial^{\mu}A^{\nu}-\partial^{\nu}A^{\mu}$ and $A^{\mu}$
is the photon field. The tree-level decay width (as function of the running
mass $x$) reads $\Gamma_{S\rightarrow\gamma\gamma}(x)=c_{\gamma\gamma}%
^{2}x^{3}/4\pi.$ Now, $c_{\gamma\gamma}$ is surely not zero because $S,$ being
made of quarks, can couple also to photons. However, its precise value is
unknown. Usually, the typical order of magnitude for $\gamma\gamma$ decays of
mesonic states amounts to some keV.

For $g\geq g_{\ast}$ there is, as described above, a (hadronic) stable state
below threshold. This state can however decay into photons. In order to
evaluate the $\gamma\gamma$ decay width the redefinition $S\rightarrow\sqrt
{Z}S$ is necessary to guarantee the correct residuum of the propagator.
Referring for example to the case $g=g_{E}$ with $z_{\text{I}}%
=(0.038-i\varepsilon)$ GeV we find for the numerical values of Eq.
(\ref{num}):
\begin{equation}
\Gamma_{E(38)\rightarrow\gamma\gamma}=Zc_{\gamma\gamma}^{2}\frac{\tilde{M}%
^{3}}{4\pi}=Z\left(  \frac{\tilde{M}}{M_{0}}\right)  ^{3}\Gamma_{S\rightarrow
\gamma\gamma}(M_{0}=1\text{ GeV})=1.55\cdot10^{-5}\text{ }\Gamma
_{S\rightarrow\gamma\gamma}(M_{0}=1\text{ GeV})\text{ .}%
\end{equation}
Assuming that $\Gamma_{S\rightarrow\gamma\gamma}(M_{0}=1$ GeV$)$ is of the
order of $10$ keV, we obtain that
\begin{equation}
\Gamma_{E(38)\rightarrow\gamma\gamma}\simeq10^{-4}\text{ keV ,}%
\end{equation}
which is a very small value. If, instead, we study the case of $M_{0}=0.45$
GeV and $g=g_{E}=2.189$ GeV we obtain $\Gamma_{E(38)\rightarrow\gamma\gamma
}=4.1\cdot10^{-4} \ \Gamma_{f_{0}(500)\rightarrow\gamma\gamma}.$ Using the
value $\Gamma_{f_{0}(500)\rightarrow\gamma\gamma}\simeq1$ keV (see the list of
results in \cite{PDG}) one gets $\Gamma_{E(38)\rightarrow\gamma\gamma}%
\simeq4.1\cdot10^{-4}$ keV, which is similar to the previous estimate.

In conclusions, we have presented a simple quantum field theoretical model
involving scalar particles and studied the poles in the first and second
Riemann sheet. We have shown that for a coupling constant large enough one
obtains, in addition to the standard resonance pole on the second sheet, also
a pole below threshold on the first sheet. The latter corresponds then to an
emergent stable state. We have studied this system using the numerical values
typical for a hadronic system and have shown that the appearance of the
additional stable state is realistic (Fig. 1 and 2). In particular, we have
studied the case in which a hadronic stable resonance with a mass of $38$ MeV
is realized: this is interesting in view of recent claims of the existence of
such a state, called $E(38)$. Indeed, it is also possible that the $E(38)$ and
the resonance $f_{0}(500)$ (alias $\sigma)$ are two manifestations of the same
object (Fig. 3). In the end, we have explored the decay of $E(38)$ into two
photons. The present theoretical description of the putative $E(38)$ state is
very different from the one put forward in\ Refs.
\cite{e38first,e38material,vanbeveren}. Thus, our analysis offers a new point
of view which may help in the process of clarification of what this state, if
existent, is or is not.

\bigskip

\textbf{Acknowledgments: }The authors thank George Rupp, Giuseppe Pagliara and
Dirk H. Rischke for valuable discussions. In particular, George Rupp is also
acknowledged for useful comments about the $E(38)$ state. T.W. thanks Dennis
D. Dietrich for helpful remarks and F.G. thanks the Foundation Polytechnical
Society Frankfurt am Main for support through an Educator fellowship.

\bigskip\appendix

\section{Next-to-leading order correction to the decay amplitude}

\begin{figure}
[ptb]
\begin{center}
\includegraphics[
height=2.1819in,
width=5.3065in
]%
{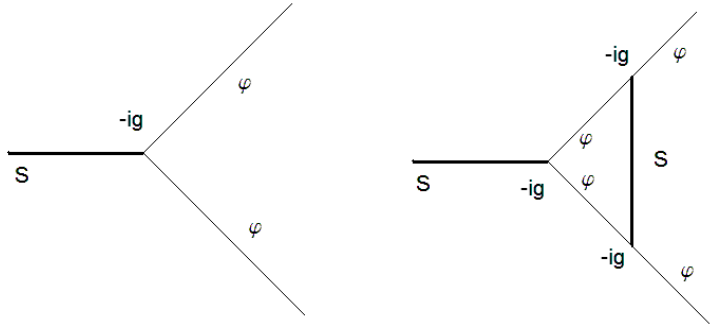}%
\caption{Tree-leve (left) and next-to-leading order (traingle diagram, right)
contributions to the decay amplitude.}%
\label{diagrams}%
\end{center}
\end{figure}

In order to establish when the coupling constant $g$ can be regarded as small
such that the resummed one-loop approximation is justified, we calculate the
(upper limit of) the next-to-leading order diagram of the amplitude of the
decay process, see Fig. \ref{diagrams}. The decay amplitude results as the sum
of two terms:%
\begin{equation}
-i\mathcal{M}=-i\mathcal{M}_{1}-i\mathcal{M}_{2}\text{ ,}%
\end{equation}
where the tree-level and the triangle contributions depicted in Fig.
\ref{diagrams} read
\begin{equation}
-i\mathcal{M}_{1}=2ig\text{ ; }-i\mathcal{M}_{2}=4g^{3}I\text{ .}
\label{amplitudes}%
\end{equation}
The coefficients $2$ and $4$ in (\ref{amplitudes}) are the appropriate
symmetry factors. The quantity $I$ is the integral%
\begin{equation}
I=I(m,M_{0})=-\int\frac{d^{4}q}{(2\pi)^{4}}\frac{\phi_{1}\phi_{2}\phi_{3}%
}{\left[  \left(  \frac{p}{2}-q\right)  ^{2}-m^{2}\right]  \left[  \left(
\frac{p}{2}+q\right)  ^{2}-m^{2}\right]  \left[  \left(  \frac{p}{2}%
+q-k_{1}\right)  ^{2}-M_{0}^{2}\right]  }\text{ .} \ ,\label{integral}%
\end{equation}
where $\phi_{i}$ is the cutoff function of the $i$-th vertex.%

We have evaluated $I$ by using the free propagator of the $S$ virtual particle
exchanged by the two $\varphi$ fields. Thus, our results represent an upper
limit for $\left\vert I\right\vert $. Namely, $\left\vert I\right\vert $ would
be actually smaller by including the full propagator of $S$, i.e. taking into
account its finite width.

We define the coupling constant $g_{\max}$ as the value of $g$ for which
$\left\vert \mathcal{M}_{1}\right\vert =\left\vert \mathcal{M}_{2}\right\vert
.$ Namely, for $g=$ $g_{\max}=1/\sqrt{2\left\vert I\right\vert }$ the
next-to-leading order contribution equals the tree-level one. Thus, for
$g>g_{\max}$ one cannot neglect the role of higher order diagrams.

In particular, we obtain the following numerical results: For $M_{0}=1$ GeV,
$g_{\max}=19.503$ GeV; the critical value $g_{\ast},$ at which the new pole in
the I-Riemann sheet emerges, reads $3.379$ and the value $g_{E},$ for which
the stable state has a mass of $38$ MeV, reads $4.877$ GeV. Both values are
safely smaller than $g_{\max}$: the amplitude ratios read $\left[  \left\vert
\mathcal{M}_{2}\right\vert /\left\vert \mathcal{M}_{1}\right\vert \right]
_{g=g_{\ast}}=0.03$ and $\left[  \left\vert \mathcal{M}_{2}\right\vert
/\left\vert \mathcal{M}_{1}\right\vert \right]  _{g=g_{\ast}}=0.06,$ thus the
tree-level diagram is by far the dominant contribution for all the range of
coupling constants relevant for our study. For $M_{0}=0.45$ GeV one gets
$g_{\max}=3.666$ GeV, which should be compared with $g_{\ast}=1.399$ GeV and
$g_{E}=2.189$ GeV. The ratios read $\left[  \left\vert \mathcal{M}%
_{2}\right\vert /\left\vert \mathcal{M}_{1}\right\vert \right]  _{g=g_{\ast}%
}=0.14$ and $\left[  \left\vert \mathcal{M}_{2}\right\vert /\left\vert
\mathcal{M}_{1}\right\vert \right]  _{g=g_{\ast}}=0.36,$ which are still
safely small, although larger than in the previous example. It should once
more be stressed that our evaluation of $\left\vert I\right\vert $ represents
only an upper limit, and the value of $g_{\max}$ is indeed larger than the
estimate given here.

All these considerations enforce our point of view that the emergence of a
pole below threshold such as the one described in this manuscript is possible
in quantum field theory. Future evaluation of higher-loop contributions is planned.

\end{document}